# MIS: Multimodal Interaction Services in a cloud perspective

[1]Patrizia Grifoni, [2]Fernando Ferri, [3]Maria Chiara Caschera,
[4]Arianna D'Ulizia, [5]Mauro Mazzei

[1, First Author]National Research Council - Institute of Research on Population and Social
Policies (CNR-IRPPS), patrizia.grifoni@irpps.cnr.it
[*2,Corresponding Author]National Research Council - Institute of Research on Population and Social
Policies (CNR-IRPPS), fernando.ferri@irpps.cnr.it
[3,4] National Research Council - Institute of Research on Population and Social Policies
(CNR-IRPPS), mc.caschera@irpps.cnr.it, arianna.dulizia@irpps.cnr.it
[5] National Research Council – mauro.mazzei@iasi.cnr.it

## Abstract

*The Web is becoming more and more a wide software framework on which each one can compose and use contents, software applications and services. It can offer adequate computational resources to manage the complexity implied by the use of the five senses when involved in human machine interaction. The core of the paper describes how SOA (Service Oriented Architecture) can support multimodal interaction by pushing the I/O processing and reasoning to the cloud, improving naturalness. The benefits of cloud computing for multimodal interaction have been identified by emphasizing the flexibility and scalability of a SOA, and its characteristics to provide a more holistic view of interaction according to the variety of situations and users.*

## 1. Introduction

Interacting in the cloud! Imagine a cloud where people and/or devices communicate and interact each other, exchanging messages in a natural manner and in an intelligent environment that is able to interpret them. This is one of the challenges of Human Machine Interaction.

"Because multimodal interaction can take place in many different hardware and software environments, it is also important for a multimodal architecture to be platform-neutral. The most obvious platform is the World Wide Web, a distributed platform where a browser runs locally on a computer or mobile device and where the application interacts with one or more remote web servers" [1].

Communication and interaction processes among humans usually involve all the five senses. The speedup of Web technologies, the availability of new devices, methodologies, theories and software for recognition of handwriting, speech, gestures, postures, gaze and facial expressions, are all contributing in the evolution of human machine interaction, involving different modalities, and considering interaction and communication in a cloud of services, addressing naturalness, in contrast with the previous generations of computerized systems, which were "sensory deprived and physically limited" as observed by Negroponte. Multimodality can improve naturalness when integrated in a cloud perspective, as it can facilitate sharing  the collective knowledge implicit for the interaction process with the environment, and for users' profiling; this knowledge provides the flexibility to make adaptable the interaction process as it allows to involve the different modalities adopting an intelligent behaviour for overcoming the artificiality of natural interaction defined by Malizia and Bellucci in [2], for the case of the gesture. This more general knowledge and vision will allow people to "interact with technology by employing the same gestures they employ to interact with objects in the real world (in everyday life)".

Since 1980 Bolt [3], a precursor of multimodal communication, with his "Put That There" describes potentialities and complexity of processes characterizing multimodality. Nineteen years later Oviatt [4] provides a systematic vision that deeply influences the design of next multimodal systems.

Currently only few of the wide range of potential inputs (gestures such as action on the touch screen, voice, facial expressions, body posture, etc.) are actually interpreted due to the complexity of parallel processes and, variety of information and data involved in managing naturalness. This requires addressing different problems at different scale levels (from the recognition of modal input, to the fusion, interpretation and disambiguation of inputs from different modalities, and the fission process) to produce multimodal outputs according to a symmetric human machine interaction.

The Web with its services can offer possible solutions, becoming a wide software framework on which each one can compose and use her/his contents, her/his software applications, and services. Since 2003 Dahl in [5] indicates Web as the platform for "future multimodal applications"; in fact W3C in 2002 launched the initiative "Multimodal Interaction Activity", aiming to support Multimodal Interaction on the Web (http://www.w3.org/TR/mmi-framework/). The Web can offer adequate computational resources to manage the complexity implied by the use of the five senses involved in human machine interaction and communication, and we assume that:

*Cloud computing and service oriented architectures provide novel perspectives for multimodal interaction, facilitating the exploitation of resources for user modeling and reasoning about context, including adapting previous cases or situations making interaction more natural.*

In fact, cloud computing and Service Oriented Architecture (SOA) -with their concepts of service, flexibility, scalability, adaptivity, evolvability, distribution, modularity, interoperability and computational power – allow to address issues underlying complex problems such as naturalness in human machine interaction.

Cloud computing is one of the emergent technology, which enables delivering IT resource as services in the Internet, and it is frequently implemented on the base of SOA architectures [6]. The cloud concept is frequently associated with "a massive network of computers and serves as a central repository of data" providing "all kinds of web services" [7]. Some main concepts of how the cloud is organized to support "the requirements of performance, scalability, reliability and availability" are also discussed in [7]. In particular, Clouds scalability, as "capability to dynamically increase or decrease the number of server instances assigned to an application to cater to the usage demands of the application" [8] allows to support the complexity of natural interaction processes using the five senses. In the last thirty years many progresses have been done for reproducing naturalness of the communication among humans in the human machine interaction and communication. Multimodal interaction is a milestone in this direction.

This perspective allows managing multimodal interaction as interchange of services, overcoming limitations connected with single platforms. This perspective will allow integrating the different resources for managing multimodal interaction available on the web with a dynamic and flexible approach, as done by Ferri et al. in [9] for other kinds o on-line resources, using the PLAKSS (Platform of Knowledge and Services Sharing) platform.

The paper is structured as follows. Section 2 provides a description of concepts, problems and approaches of multimodal interaction; section 3 introduces the main concepts on SOA and their implications when considering interaction and communication as virtualized services, and on Cloud Computing in particular.

Section 4 provides the cloud perspective of multimodal interaction. Finally section 5 concludes the paper.

## 2. The challenge Multimodal Interaction

There is a wide scientific literature that considers human machine interaction as a process involving the five senses: it is the literature on multimodal interaction methods and systems. It addresses a multiplicity of problems, such as combining visual information with voice, gestures and other modalities, and provides flexible and powerful dialogue approaches.

General concepts and an architecture supporting multimodal interaction were proposed by Oviatt in [10]. This architecture consists in the acquisition, recognition, integration and decision levels. The acquisition level is the interface with which users put their input by speech, sketch, handwriting, and so on. The different inputs are processed by the recognition level and they are integrated by a multimodal fusion system, which merging and synchronizing information from the different modalities produces a meaningful and correct input. The decision level, with its dialogue management system, processes the integrated multimodal message/command by activating required applications and services. Other authors proposed architectural schema providing fusion at recognition level and, at decision level. The integration of multimodal input signals at recognition level is based on the use of appropriate structures such as action frame [11], input vectors [12] and slots [13]. In the decision-based approach, the dialogue management system integrates the output of each modal recognizer using dialogue-driven fusion procedures.

D'Ulizia et al. [14] [15] proposed a linguistic approach structuring all features of multimodal communication in a multimodal language based on a multimodal grammar, defined in an easy and intuitive way using a "by example" paradigm. This paradigm can be also adopted for a more natural interaction in mechatronic systems, as defined by Ferri et al. in [16]. This approach, supported by the use of algorithms learning and evolving linguistic abilities of the system [17] represents a step toward naturalness.

Naturalness in multimodality can simplify the user's activity and reduce cognitive load; but in this case systems have to manage problems of the human-human communication, such as ambiguity. An ambiguity arises when humans communicate using their five senses and more than one interpretation is possible. Ambiguities are active elements in group communication processes and they are frequently used in an active manner to negotiate an intended meaning, to identify problems and their solutions and, afterwards, to refine them [18][19][20] making processes more flexible and natural.

In the literature there exist different approaches proposed for managing ambiguities; for example, Lee et al. [21] as well as Avola et al. [22][23] focus on the problem of single modal ambiguity; Mankoff, and Abowd, in [24] continue to focus themselves on modal ambiguities, and use multimodality for addressing a user mediated approach for the disambiguation process. Caschera et al., in their papers [25][26][27] do the shift, discussing of ambiguities also in the case of combined multimodal inputs, focusing on ambiguities produced both by the propagation at a multimodal level of ambiguities arising by the recognition process of single modality, and by the combination of unambiguous modal information generating contrasting concepts at a multimodal level. In particular, Caschera et al. in [26] provide an approach based on Hidden Markov Models to solve multimodal ambiguities, shifting on the system side the load of producing one only interpretation. In this situation the use of integrated multiple input modes can enable users to benefit from the natural communication among humans. An important for multimodal interaction consists of the arrangement of outputs from the different modalities and this process is identified as multimodal fission. An overview on the problem and some solutions proposed in literature is provided in [28]

An example of multimodal system for mobile devices supporting human-like interactions between the human and the device was developed by Schuricht et al. [29].

A shift toward communication and interaction, considered as services in a cloud of services and resources, is now necessary. In fact, small and mobile devices such as smart phones and sensors are able to capture, manage and return signals by different communication channels.

Beyond the capability to capture the different signals from humans, there are devices able to capture signals from the environment where they are situated; therefore they can acquire contextual information such as location (by GSM), environment (by camera), environmental noises (by microphone), etc.. Different devices exchange signals, interacting and communicating each other's, with sensors distributed in the environment and with people by different communication technologies such as Bluetooth, infrared, etc, allowing to acquire relevant information to know and better understand the context and events characterizing it and addressing naturalness.

*Interaction and communication are virtualized services on the Web as part of a global system*

Naturalness, complexity of communication and interaction processes imply controlling and managing all processes connected with interpretation, ambiguity management, users and context adaptivity, and the language evolutionary nature. These issues require optimizing resources involved in these processes respect to their naturalness.

Capturing knowledge, information and services, providing and managing interaction according to this complex perspective, when sensors and mobile devices are involved in, requires that each sensor and device will be considered as part of a global system with a distributed computational capability.

All these features need to be located in the perspective of services; in particular services involved in the communication process can use concepts and resources of SOA; Cloud Computing and users do not need to have knowledge of, expertise in, or control over the technology infrastructure in the "Cloud" that supports them. Before providing this perspective, it is necessary to introduce the main concepts on SOA in general and implied in considering interaction and communication as virtualized services, and on Cloud Computing in particular.

## 3. SOA, Web service and Cloud computing

A services system is defined in [30] as *"a dynamic configuration of resources (people, technology, organizations and shared information) that creates and delivers value between the provider and the customer through service. In many cases, a service system is a complex system in that configurations of resources interact in a non-linear way. Primary interactions take place at the interface between the provider and the customer. However, with the advent of ICT, customer-to-customer and supplier-to-supplier interactions have also become prevalent. These complex interactions create a system whose behavior is difficult to explain and predict"*.

In the SOA manifesto [31] there is a shift from the concept of service system to the more general concept of service orientation, defined as "a paradigm that frames what you do". The concept was then directed towards the architectural perspective with Sun, which defined SOA, in the late 1990's to describe Jini, an environment for dynamic discovery and use of services over a network.

Sometimes, SOA and Web services are used for addressing the same meaning. SOA approach can be implemented according to different strategies and one of most frequently used refers to Web services. W3C has defined Web services starting from the concept of service introduced in SOA, defining them as "software systems designed to support interoperable machine to machine interaction over a network".

WSDL is the XML format describing Web services interfaces. Other systems interact with Web services using messages enclosed in a SOAP envelope; they are transferred using Hypertext Transfer Protocol (HTTP) along with XML and other Web-related standards. This approach guarantees interoperability, as it provides independence from the platform and from the used language.

SOA and Web services with their different features can provide the framework for structuring the complexity of human machine interaction processes in a cloud perspective.

*Cloud computing represents an opportunity for getting up accessible information and resources on the Internet and building applications leveraging them.*

There are many ways in which people define cloud computing, but some relevant features identified by Devi [32] common to all definitions are:

- *"Dynamic scalability on demand: Cloud computing is expected to provide a computing capability that can scale up (to massive proportions) or scale down dynamically based on demand.*
- *Virtualized resources: Computing resources offered in cloud computing are virtualized, meaning they make it unnecessary for consumer of the resources to worry about details of the underlying layer. The way virtualized resources are organized can be very different from the actual organization of the physical resources.*
- *Resources are provided as a Service: Service can include anything that is useful and can be effectively delivered to a consumer asking for it"*.

Cloud architectures are based on three main concepts (and related levels: 1) Software as a Service (SaaS), 2) Platform as a Service (PaaS) and 3) Infrastructure as a Service (IaaS).

In the SaaS, the consumer uses the provider's applications running on a public Cloud infrastructure. For example s/he rather than buying an application to run on her/his in-house systems, s/he can use the application provided as a service on the Cloud. In the PaaS the provider supplies the developing environment (from the operating systems, middleware and programming tools) allowing the consumer to develop applications onto this platform. Differently from PaaS, with IaaS, the consumer deploys and runs arbitrary software, which can include operating systems and applications, on the provider's infrastructure.

These different kinds of cloud services correspond to an SOA system including software services, an SOA platform and a supporting infrastructure. That is also true for services supporting multimodal interaction.

The question is how to create a Cloud of services. This question implies the need to virtualize services and to manage the complexity of this virtualization. The cloud computing proposed solution is defined through SOA.

*Each thing (e.g. functions managing multimodal interaction) can be conceived as a service.*

In this perspective, also the human machine interaction and communication processes will be designed and implemented considering them as part of the cloud and human machine interaction processes can be modeled by SOA technologies as each HCI function can be considered as a service.

## 4. Multimodal Interaction in the cloud perspective

Interacting in a natural way using the five human senses with mobile devices and/or embedded sensors distributed in the environment requires managing multimodal interaction in a distributed, flexible, modular, interoperable and scalable manner; Cloud Computing and SOA can address these requirements.

Naturalness of communication and interaction processes imply a high complexity for managing information broken down and stored across a network of services acting as a cloud; similarly human brain manages communication and interaction processes between humans using dynamic scalability in assigning brain resources according to its specific functions; functioning and shape of cloud bring to mind respectively the human brain functioning and shape (see Figure 1).

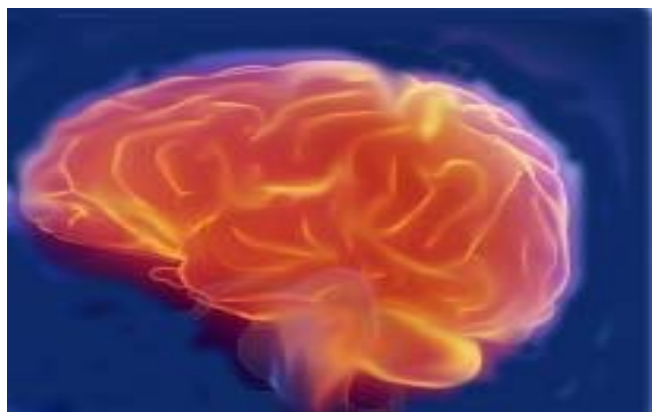

**Figure 1.** Cloud or brain?

Multimodal interaction can be therefore managed in a cloud perspective as a set of services: from the I/O communication management (I/OCM) to the interaction broker given by a Brokerage Service (BS), mainly devoted to coordinate the different hardware and software resources involved in the I/O processes, the recognition of the different inputs, a service for their combination (fusion), interpretation and disambiguation, a service for arranging the different outputs (fission) [*]. All services are arranged in a cloud implementing naturalness of interaction with their necessary dynamic scalability, using their knowledge on the different users and contexts and services such as user profile.

The I/OCM service - provided at SaaS level - manages the I/O of the Web interfaces from/to users, sensors and devices controlling the different resources involved in adapting the interface to the device, to the user and to the domain characteristics (at PaaS level) in order to guarantee naturalness of the interaction process. In particular, the I/OCM requires and uses services for defining and managing the different users profiles and contexts modeling (e.g. services for Defining and Managing Agents – DMA - and, services for building knowledge organizing them in ontologies using an Ontology Editing service – OE, as figure 2 shows).

Naturalness implies intelligence of interaction processes, and a Rule based Inference Engine service (RIE) can be used to create and manage knowledge formalized with the defined ontology providing the structure of an intelligent behavior.

As it happens in the human brain for the communication process, which develops a specialization for language, also the human machine interaction process consists of parts that can be associated with interaction services, in the holistic perspective of the linguistic approach implied by the concept of multimodal language.

A service is devoted to define the Multimodal Language and its Grammar (DMLG). The Recognition, Interpretation/disambiguation Services (RS and IS), as well as the FuSion and FiSsion (FuS, FiS) are also defined. These services are all organized at PaaS level. Information, knowledge and data related with users, contexts, interaction language, ontologies, are all provided at IaaS level.

This cloud of services needs to be organized and managed in order to optimize the use of the different resources making the interaction natural and coherent with the users and context features. For this reason a Brokerage Service (BS) is foreseen, assigning the different resources involved in the recognition, interpretation/disambiguation, fusion and fission processes. In fact the I/CMO, once received input, sends signals to the modal recognizers and then, it decides which of these will contribute to the process of fusion, definition and management of user profiles, creation and evolution of languages, of fission, and so on. The execution of these processes (fusion, fission, Definition and Management Agents, ...) requires resources; their optimal management is guaranteed by the Broker (BS).

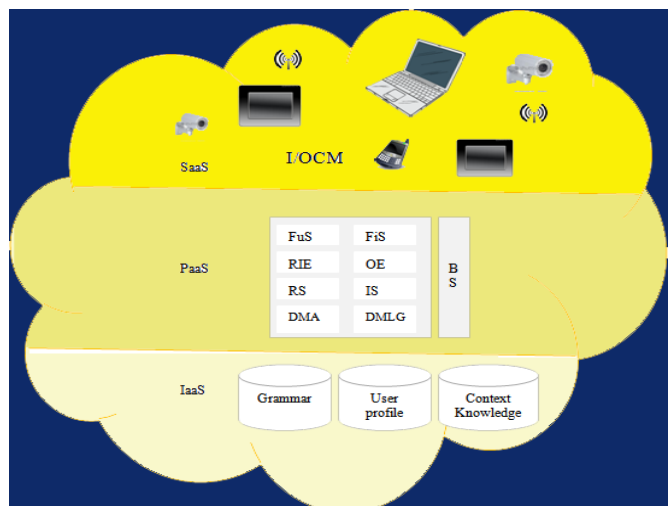

**Figure 2.** Multimodal Interaction in the Cloud

In this scenario each thing is a service, and also the multimodal interaction management is considered according to this perspective.

## 4.1. SOA perspective

Each service supports actions of the interaction process; the data service messaging, their management and organization are consumed via World Wide Web.

The interaction events on the client side activate the communication process; the different services are associated by the orchestration process to the recognition of the different modalities, their interaction management, fusion, fission and interpretation/disambiguation processes. These services involve data and metadata for managing multimodal interaction defining, according to a linguistic approach, Multimodal Sentences and Multimodal Languages.

The main building blocks, according to an SOA perspective, are the service provider, the service registry, and the service consumer. The provider publishes a description of services given via a service registry. The service consumers consume (as a resource) the service. Clients must be able to find the description of services that they require, and must be able to bind to them.

Multimodal interaction and the set of services designed and organized according to a Service Oriented Architecture are proposed in Figure 3.

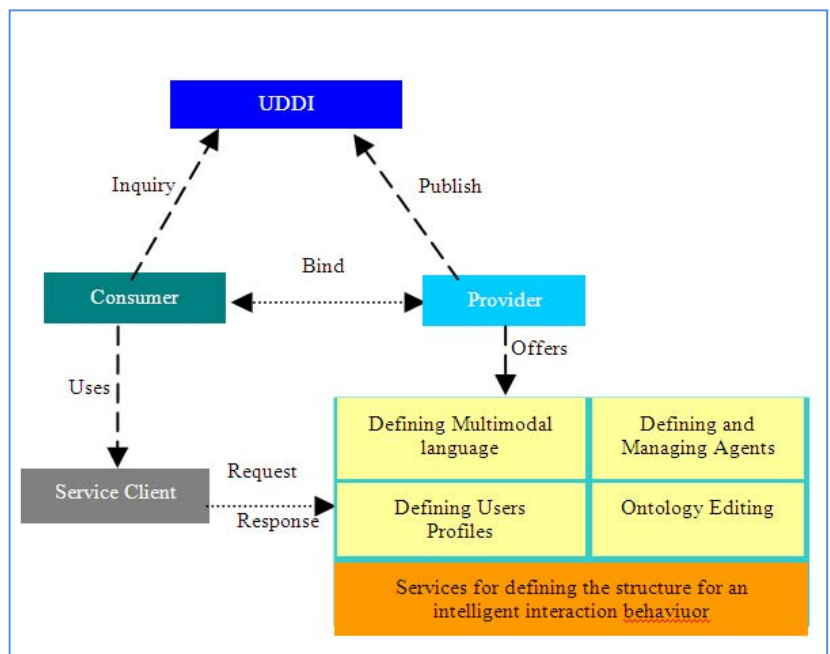

**Figure 3.** Multimodal Interaction: a SOA perspective

The service consumer searches for the registry where finding services, then the consumer can invoke the operations for the service.

Combining multimodal inputs, the execution of services for interpreting them and services for producing multimodal outputs, are all dynamic issues. In particular, when a user sends one or more modal input for a given time interval, the service consumer, on the client side requires one or more than one service for the modal recognition by the Multimodal Interaction Management service. The service provider offers services publishing them in the UDDI register.

The offered services are hierarchically structured. In particular, a parent/child association exists between the I/OCM and the other services.

In particular, inputs on the client side use Web services to invoke the associated service; an XML file is sent to the modality recognition service for each input channel (speech, sketch and each other modality involved in the input). The I/OCM requires Web services to recognize the different modal inputs.

When the different modal input recognizers return their responses to the Multimodal Interaction Management service (MIMs), it send its request to the multimodal fusion service, which returns the multimodal sentence associated with the multimodal input to the MIMs.

MIMs requires the interpretation/disambiguation service for the Multimodal Sentence on the base of data and metadata associated with the syntax, the semantic and the pragmatic of the multimodal language. The MIMs, on the base of the interpretation/disambiguation, requires a response by the system (a multimodal output) that is produced by the multimodal Fission Service (FiS).

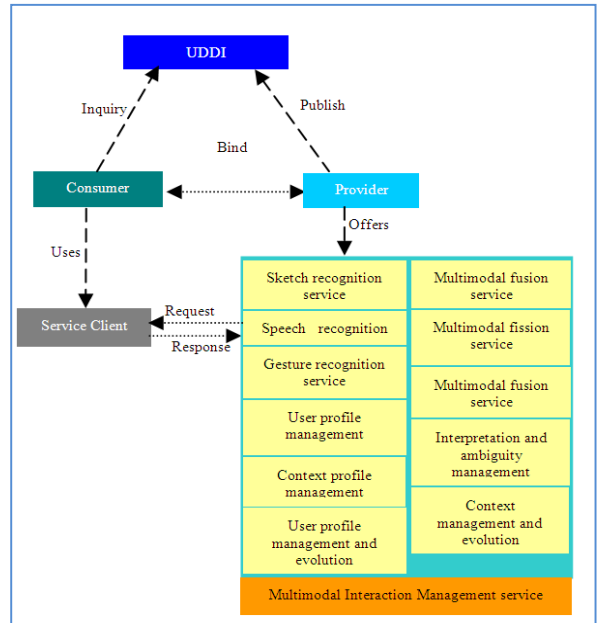

**Figure 4.** Services for defining and managing knowledge

Are also provided, (see Figure 4) services for Defining Multimodal Languages (DML), and services for defining and managing knowledge and in particular for: Defining and Managing Agents (DMAG) for Defining Users Profiles (DUP), for Ontology Editing (OE) related with the application domain and with the context for providing the structure for an intelligent and natural interaction behavior.

## 5. Conclusion

This article addresses perspectives of a natural human machine interaction based on Cloud and SOA concepts. This perspective allows managing complex processes also using thin clients, currently very widely used by people.

Cloud computing is a conceptualization of delivered and shared scalable, configurable computing resources that can be dynamically and automatically provisioned and released. Various users sharing informational resources extended in space and time with no distinctly specified software/hardware boundaries and having a collective behavior can be part of the cloud by mobile devices, sensors and wireless devices. The location of resources is in the "cloud" of thousands of computerized devices linked together. Each service for the multimodal interaction cooperates in the multimodal communication process. A crucial issue will be represented by the fact that "clients only pay for the quantity of the rented resources (data storage, computation, etc.) they consume", but cooperation approaches [33] allowing to overcome this kind of problem are emerging.Public and private information, related with data of different kinds, human preferences and behaviors related with a specific context, available services for a specific context are all elements available by a wide and heterogeneous people. The majority of computing will be more and more in the cloud in the next future and they will be based on heterogeneous mobile and wireless devices. In this scenario cloud can provide the necessary scalability and flexibility for providing the holistic approach devoted to improve naturalness of interaction. This phenomenon is changing the users' needs in terms of interaction paradigm and the focal issue discussed in this article is:

*What new kind of interaction will help people in accessing/ sharing massive amounts of data in different situations and using different devices in a natural manner?*

It is our opinion that new interaction will represent the confluence of research for more natural interaction in the cloud. Naturalness involves complex processes and the cloud allows to manage this complexity organizing and providing the necessary services and resources.

## 6. References

[1] Dahl, D., "The W3C Multimodal Architecture and Interfaces Standard", Journal of Multimodal User Interfaces, Springer, vol. 7, no 3, pp. 171-182, 2013.

[2] Malizia,A., Bellucci, A:, "The artificiality of natural user interfaces", Commun. ACM, ACM, vol. 55,. No 3, pp. 36-38, 2012.

[3] Bolt, R.A. "Put that there: Voice and gesture at the graphics interface" ACM Computer Graphics, ACM, vol. 14, no 3, pp. 262–270, 1980.

[4] Oviatt, S., "Ten Myths of Multimodal Interaction", Commun. ACM, ACM, vol. 42, no 11, pp. 74-81, 1999.

[5] Dahl, D., "Interview by Paolo Baggia (Loquendo)", 2003. http://www.vxmlitalia.com/dahl_eng.html

[6] Zhang, H., Yang, X., "Cloud Computing Architecture Based-On SOA", In Proceedings of the 2012 Fifth International Symposium on Computational Intelligence and Design (ISCID '12), pp. 369-373, 2012.

[7] Kim, W., "Cloud computing architecture", International Journal of Web and Grid Services, Inderscience, vol.9, no 3, pp. 287-303, 2013.
Link - http://inderscience.metapress.com/content/C10011247524MK45

[8] Wu, J., Liang, Q., and Bertino, E., "Improving Scalability of Software Cloud for Composite Web Service", In Proceedings of the IEEE International Conference on Cloud Computing, pp. 143-146, 2009.

[9] F Ferri, P Grifoni, MC Caschera, A D'Ulizia, C Praticò, "KRC: KnowInG crowdsourcing platform supporting creativity and innovation", AISS: Advances in Information Sciences and Service Sciences, AICIT, vol. 5, no 16, pp. 1-15, 2013.

[10] Oviatt, S. L. "Multimodal interfaces", in Handbook of Human-Computer Interaction, ed. by J. Jacko & A. Sears, Lawrence Erlbaum: New Jersey, 2002.

[11] Vo M.T. "A framework and Toolkit for the Construction of Multimodal Learning Interfaces", PhD. Thesis, Carnegie Mellon University, Pittsburgh, USA, 1998.

[12] Pavlovic, V.I., Berry, G.A., Huang, T.S. "Integration of audio/visual information for use in human-computer intelligent interaction", In Proceedings of the IEEE International Conference on Image Processing (ICIP '97), pp. 121-124, 1997.

[13] Andre, M., Popescu, V.G., Shaikh, A., Medl, A., Marsic, I., Kulikowski, C., Flanagan J.L. "Integration of Speech and Gesture for Multimodal Human-Computer Interaction", In Second International Conference on Cooperative Multimodal Communication, pp. 28-30, 1998.

[14] D'Ulizia, A., Ferri, F., and Grifoni, P., "Generating Multimodal Grammars for Multimodal Dialogue Processing", . IEEE Transactions on Systems, Man, and Cybernetics, Part A, IEEE, vol. 40, no 6, pp. 1130-1145, 2010.

[15] D'Ulizia A, Ferri F., Grifoni P.: A Hybrid Grammar-Based Approach to Multimodal Languages Specification, OTM 2007 Workshop Proceedings, 25-30 November 2007, Vilamoura, Portugal, Springer-Verlag, Lecture Notes in Computer Science 4805, pp 367-376, 2007.

[16] Ferri, F., D'Ulizia, A., and Grifoni, P., "Multimodal Language Specification for Human Adaptive Mechatronics", JNIT: Journal of Next Generation Information Technology, AICIT, vol. 3, no 1, pp. 47 – 57, 2012.

[17] D'Ulizia, A., Ferri, F., and Grifoni, P., "A Learning Algorithm for Multimodal Grammar Inference", IEEE Transactions on Systems, Man, and Cybernetics, Part B, IEEE, vol. 41, no 6, pp. 1495-1510, 2011.

[18] McLuhan, M. and Fiore, Q., "The Medium Is the Massage", New York: Random House, 1967.

[19] Stacey, M. and Eckert, C. "Against Ambiguity" (CSCW) Computer Supported Cooperative Work, Kluwer Academic, vol.12, no 2, pp. 153–183, 2003.

[20] Aoki P. M. and Woodruff, A. "Making Space for Stories: Ambiguity in the Design of Personal Communication Systems" In Proceedings of the ACM Conference on Human Factors in Computing Systems (CHI 2005), pp.181-190, 2005.

[21] Lee, K.-B., Jin, S.-H., Hong, K.-S. "An Implementation of Multimodal User Interface using Speech, Image and EOG", IJEI: International Journal of Engineering and Industries, AICIT, vol. 2, no2, pp.76 -87, 2011.

[22] Avola, D., Caschera M. C., Ferri F., Grifoni P., "Ambiguities in Sketch-Based Interfaces", In: The Proceedings of the 40th Hawaii International Conference on System Sciences (HICSS 2007), IEEE Press, p. 290b. 2007.

[23] Avola D., Caschera M. C., Ferri F., Grifoni P. "Classifying and resolving ambiguities in sketch-based interaction". International Journal of Virtual Technology and Multimedia, Inderscience , vol. 1, no 2, pp. 104-139, 2010.

[24] Mankoff, J. and Abowd, G., "Error Correction Techniques for Handwriting, Speech, and Other Ambiguous or Error Prone Systems", Georgia Tech GVU Center Technical Report, GIT-GVU-99-18, 1999.

[25] Caschera M.C., Ferri F., Grifoni P. "An Approach for Managing Ambiguities in Multimodal Interaction", OTM 2007 Ws, Part I, LNCS 4805. Springer-Verlag Berlin Heidelberg , pp. 387–397, 2007.

[26] Caschera, M.C., Ferri, F., Grifoni, P., "From Modal to Multimodal Ambiguities: a Classification Approach", JNIT: Journal of Next Generation Information Technology, AICIT, vol. 4, no 5, pp. 87 – 109, 2013.

[27] Caschera, M.C., Ferri, F., Grifoni, P. „InteSe: An Integrated Model for Resolving Ambiguities in Multimodal Sentences", IEEE Transactions on Systems, Man, and Cybernetics: Systems, IEEE, vol. 43, no 4,, pp. 911 – 931, 2013.

[28] Grifoni, P., "Multimodal Fission" In: P. Grifoni, Ed., Multimodal Human Computer Interaction and Pervasive Services, IGI Global, pp. 103, 120, 2009.

[29] Schuricht, M., Davis Z., Hu, M., Prasad, S., Melliar-Smith, P.M. & Moser, L.E., "Managing multiple speech-enabled applications in a mobile handheld device", International Journal of Pervasive Computing and Communications, Emerald group, vol. 5, no 3, pp. 332–359, 2009.

[30] IfM and IBM. "Succeeding through Service Innovation: A Service Perspective for Education, Research, Business and Government", Cambridge, United Kingdom: University of Cambridge Institute for Manufacturing,2008.

[31] Arsanjani, A., Booch, G., Boubez, T., Brown, P. C., Chappell, D., deVadoss, J, Erl, T., Josuttis, N., Krafzig, D., Little, M., Loesgen, B., Manes, A. T., McKendrick, J., Ross-Talbot, S., Tilkov, S., Utschig-Utschig, C., Wilhelmsen, H., 2009,http://www.soa-manifesto.org/

[32] Devi, Y 2009. http://www.siliconindia.com/guestcontributor/guestarticle/175/Exploring_Synergy_between_SOA_Cloud_Computing_Yogesh_Devi.html

[33] Mousannif, H., Khalil, I., and Kotsis, G. "The cloud is not 'there', we are the cloud!", International Journal of Web and Grid Services, Inderscience , vol. 9, no 1, pp. 1-17, 2013.